\documentstyle[]{article}
\font\cero=cmss10 scaled 1728 \font\uno=cmssbx10 scaled 1200
\begin{document}
\small{
\begin{flushleft}
{\cero Basic symplectic geometry for p-branes with thickness in a curved background} \\[3em]
\end{flushleft}
{\sf Alberto Escalante}\\
{\it Instituto de F\'{\i}sica, Universidad Aut\'onoma de Puebla,
Apartado postal J-48 72570, Puebla Pue., M\'exico.}
(aescalan@sirio.ifuap.buap.mx) \\[4em]

\noindent{\uno Abstract} \vspace{.5cm}\\ We show that the Witten
covariant phase space for p-branes with thickness in an arbitrary
background is endowed of a symplectic potential, which although is
not important to the dynamics of the system, plays a relevant role
on the phase space, allowing us to generate a symplectic structure
for the theory and therefore give a covariant description
of canonical formalism for quantization.  \\

\noindent \\

\begin{center}
{\uno I. INTRODUCTION}
\end{center}
\vspace{1em} \ As we know, a covariant description of the
canonical formalism for quantization and the study of the symmetry
aspects has been given by using basic ideas of symplectic
geometry. This formalism, has all virtues that Feyman's path
integral has, that is, manifestly covariant, maintaining all
relevant symmetries, such as Poincar\'e invariance. This scheme,
also has been applied in many theories, for example, Witten {\it
et al}
 take the cases of Yang-Mills and General Relativity \cite{1}, open superstrings \cite{2},
 and the analysis of diffeomorphism
invariant field theories was considered by Wald {\it et al}
\cite{3}, among others. Recently this formalism was taken up by
Cartas-Fuentevilla to p-branes governed by the Dirac-Nambu-Goto
action [DNG] \cite{4}, using a weakly covariant formalism for
deformations, introduced by Capovilla-Guven [CG] \cite{5}, and in
\cite{6} using
a strongly covariant formalism, introduced by Carter \cite{7}.\\
On the other hand, in many cases it was seen that [DNG] action is
inadequate and there are missing corrective quadratic terms in the
extrinsic curvature. For example, in the eighties Polyakov
proposed a modification to the [DNG] action by adding a rigidity
term constructed with the extrinsic curvature of the worldsheet
generated by a string, and  to include quadratic terms in the
extrinsic curvature to the [DNG] action is absolutely necessary,
because of its influence in the infrared region determines the
phase structure of the string theory, in this manner, we can
compute the critical behavior of random surfaces an their
geometrical and physical characteristics \cite{8}. In the
treatment of topological defects \cite{9}, curvature terms are
induced by considering an expansion in the thickness of the
defect. Bosseau and Letelier have studied cosmic strings with
arbitrary curvature corrections, finding for example, that the
curvature correction may change the relation between the string
energy density and the tension \cite{10}. Furthermore, such models
have been used to describe mechanical properties of lipid
membranes \cite{11}. More recently, conservation laws for bosonic
brane dynamics have been obtained for an action quadratic
in the extrinsic curvature \cite{12}. \\
Due to the above ideas, the purpose of this article is to
establish the bases of the covariant canonical formalism for
corrections to the [DNG] action, which
depend quadratically on extrinsic curvature ( we will denote this corrective term by [QEC]).\\
This paper is organized as follows. In Sect.II, we start with the
formalism of deformations introduced by [CG] [7], and we give some
remarks for [DNG] p-branes, obtaining by another way the results
obtained by Cartas-Fuentevilla \cite{4}. In Sect.III, from
tangential deformations, we obtain a symplectic potential for
[QEC] action, and the linearized equations of motion taking as
special case a extremal surface in an arbitrary background which
will be useful in the next section. In Sect.IV, we define the
Witten covariant space phase for [QEC] theory, and considering the
linearized equations for [QEC] action, we obtain a covariant
conserved current by applying the self-adjoin operators method. In
Sect.V, we find a two-form for [QEC] theory and we show that is an
exact and no-degenerate differential form, from the global
potential, found in the Sect. III. In Sect. VI we establish some
remarks and
prospects.\\

\setcounter{equation}{0} \label{c2}.

\noindent \textbf{II. Global sympletic potential for Dirac-Nambu-Goto action}\\[1ex]
In \cite{6}, Cartas-Fuentevilla showed using a strongly covariant
formalism, that the [DNG] action has a covariant conserved
symplectic current obtained from a global symplectic potential. In
the same way, we shall show that in the weakly covariant formalism
used in the present treatment \cite{5} there exists also a global
symplectic potential for [DNG] action, from which we will get by
another way the symplectic current obtained by
Cartas-Fuentevilla  in \cite{4}. \\
To prove it, we take the [DNG] action, that is proportional to the
area of the spacetime trajectory created by the brane,
\begin{equation}
S=-\mu \int \sqrt{-\gamma}d^{D} \xi,
\end{equation}
where $\mu$ is the brane tension. In agreement with \cite{5}, we
take the tangential and normal deformations of the action (1), and
we obtain
\begin{equation}
\delta S= - \mu \int \sqrt{-\gamma} \nabla_{a} \Phi^{a} d^{D} \xi
- \mu  \int K^{i} \phi_{i}d^{D} \xi,
\end{equation}
where
\begin{equation}
\phi^{a}= \delta X^{\mu}e{^{a}}_{\mu} \quad {\rm{and}} \quad
\phi^{i}= \delta X^{\mu} n{^{i}}_{\mu},
\end{equation}
$\delta X^{\mu}$ being the infinitesimal spacetime variation of
the embbeding, with $n^{i}$ and $e^{a}$ as the vector fields
normal and tangent to the
worldsheet respectively.\\
We can see that the second term on the right hand-side of equation
(2) corresponds to the equation of motion, $K^{i}=0$, and the
corresponding linearized equations are \cite{4}
\begin{equation}
[\widetilde
\Delta^{i}_{j}+K{_{ab}}^{i}K{^{ab}}_{j}-g(R(e_{a},n_{j})e^{a},n^{i})
] \phi^{j}=0,
\end{equation}
where $\widetilde \Delta= \widetilde \nabla^{a} \widetilde
\nabla_{a}$, $K{_{ab}}^{i}$ is the extrinsic curvature, and
$g(R(e_{a},n_{j})e^{a},n^{i})= R_{\alpha \beta \mu \nu}
n{_{j}}^{\alpha} e{_{a}}^{\beta} e{^{a}}^{\mu} n{^{i}}^{\nu}$,
being $R_{\alpha \beta \mu \nu}$ the background Riemann tensor (for more detail see the Appendix).\\
 On the other hand, the argument of the total divergence,  $\phi^{a}$, given in (2), are
 neglected in the literature
 because of is not relevant locally to the dynamics of the system.
However, we will identify $-\sqrt{-\gamma}\phi^{a}$ from the first
term on the right-hand side in equation (2) as a symplectic
potential on the phase space and we will take its variation (its
exterior derivative on $Z$, see the Appendix, equation (55) ),
this is
\begin{equation}
D_{\delta}(-\sqrt{-\gamma}\phi^{a})=\sqrt{-\gamma}[K^{abi}
\phi_{i} \phi_{b} + \phi_{i} \widetilde \nabla^{a}\phi^{i} ].
\end{equation}
 It is important
to notice that because of $\phi^{a}$ is a diffeomorphism on the
world-sheet, it can be gauged away in the equation (5), although
its variation does not vanish, thus
\begin{equation}
D_{\delta}(-\sqrt{-\gamma}\phi^{a})=\sqrt{-\gamma}[ \phi_{i}
\widetilde \nabla^{a}\phi^{i}],
\end{equation}
in this manner,  we can see that the last equation is the
symplectic current obtained by Cartas-Fuentevilla \cite{4}
applying the self-adjoint operators method. Thus, equation (6)
implies that the symplectic structure obtain in \cite{4} is no
only a closed two-form but even an exact two-form.\\
Therefore, we can identify indeed $-\sqrt{-\gamma}\phi^{a}$ as a
global symplectic potential for [DNG] p-branes, that can not be
neglected because of allow us construct geometrical structures
physically relevant on the phase space $Z$. Following these ideas,
in the next section we shall consider an action quadratic in the
extrinsic curvature.
\newline
\newline
\noindent \textbf{III.The quadratic term in the extrinsic curvature }\\[1ex]
Let us consider the following action quadratic in the extrinsic curvature
\begin{equation}
S_{2}=\alpha \int d^{D}\xi \sqrt {-\gamma} K_{i}K^{i},
\end{equation}
where
\begin{equation}
K^{i}= \gamma^{ab}K{_{ab}}^{i},
\end{equation}
and $\alpha $ is the brane tension coefficient. As in the last
section, and using the [CG] deformation formalism \cite{5}, we
take the tangential and normal deformations of the world-volume
and we obtain,
\begin{eqnarray}
\nonumber \delta{S_{2}} \!\!\! & = & \!\!\! 2\alpha \int d^{D}\xi \sqrt {-\gamma}\left[-\widetilde \triangle K^{i}\phi_{i} + \left(g(R(e_{a},n^{j})e^{a},n^{i}) - (\gamma ^{ac}\gamma ^{bd} - \frac{1}{2}\gamma ^{ab}\gamma ^{cd} )K{_{ab}}^{j}K{_{cd}}^{i}\right)\phi_{i} K_{j} \right] \\
\!\!\! & + & \!\!\! 2\alpha \int d^{D}\xi \nabla_{a}\left[ \sqrt
{-\gamma}\left(\frac{1}{2}K^{j}K_{j} \Phi^{a} + \phi_{i}\widetilde
\nabla^{a}K^{i}-K_{i}\widetilde \nabla^{a}\phi^{i} \right)\right],
\end{eqnarray}
where we can find the equations of motion
\begin{equation}
\widetilde \triangle K^{i}+
\left(-g(R(e_{a},n^{j})e^{a},n^{i})+(\gamma ^{ac}\gamma
^{bd}-\frac{1}{2}\gamma ^{ab}\gamma ^{cd}
)K{_{ab}}^{j}K{_{cd}}^{i}\right)K_{j}=0,
\end{equation}
and we identify from the pure divergence term in (9), the
following
\begin{equation}
\Psi^{a}= \sqrt {-\gamma} \left[\frac{1}{2}K^{j}K_{j} \phi^{a} +
\phi_{i} \widetilde \nabla^{a}K^{i}- K_{i}\widetilde
\nabla^{a}\phi^{i}\right],
\end{equation}
 as a  symplectic
potential for [QEC] p-branes, that are neglected, because of does
not contribute locally to the dynamics of system, but it generates
a geometrical structure
on the phase space, as we will see in the next section.\\
 We
can obtain the linearized equations taking the variation of the
equation (10), which are:
\begin{eqnarray}
\nonumber \!\!\! & - & \!\!\! \widetilde \Delta \widetilde \Delta
\phi^{i}- 2K{^{ab}}_{j}K^{j}( \widetilde \nabla _{a}\widetilde
\nabla _{b} \phi^{i})+ \frac{1}{2} K^{j}K_{j}\widetilde \Delta
\phi ^{i}+ (K^{i}K_{j}-2K{_{ab}}^{i}K{^{ab}}_{j}) \widetilde
\Delta \phi ^{j}
\\ \nonumber
 \!\!\! & - & \!\!\!  2K{^{ab}}_{j}(\widetilde \nabla _{a}K^{j})(\widetilde \nabla _{b}\phi ^{i})- K_{j}(\widetilde\nabla _{a}K^{abj})(\widetilde \nabla _{b}\phi ^{i}) -2K{^{ab}}_{j}(\widetilde \nabla
_{a}K^{i})(\widetilde \nabla _{b}\phi ^{j})\\ \nonumber \!\!\! & -
& \!\!\! 2 \widetilde
\nabla^{c}[K{_{ab}}^{i}K{^{ab}}_{j}](\widetilde \nabla _{c}\phi
^{j})+2K^{abi}(\widetilde \nabla _{a}K_{j})(\widetilde \nabla
_{b}\phi ^{j}) - \widetilde \Delta [K{_{ab}}^{i}K{^{ab}}_{j}]\phi
^{j} \\ \nonumber \!\!\! & + & \!\!\! K_{j}(\widetilde \nabla
^{b}K^{i})(\widetilde \nabla _{b}\phi ^{j})+ K_{j}(\widetilde
\nabla _{a}K^{abi})(\widetilde \nabla _{b}\phi ^{j}) + \widetilde
\nabla_{a}K_{j} \widetilde \nabla^{a}K^{i} \phi^{j} -2(\widetilde
\nabla
_{b}K^{i})(\widetilde \nabla _{a}K{^{ab}}_{j})\phi ^{j} \\
\nonumber
\!\!\! & - & \!\!\! 2K{^{ab}}_{j}(\widetilde \nabla _{a}\widetilde \nabla _{b}K^{i})\phi ^{j}+2K{_{ab}}^{i}K{^{bc}}_{k}K^{a}{_{cj}}K^{j}\phi ^{k}+ \frac{1}{2}K{_{ab}}^{i}K{^{ab}}_{k}K^{j}K_{j}\phi ^{k}+K^{i}K_{j}K{_{ab}}^{j}K{^{ab}}_{k}\phi ^{k} \\
\!\!\! & - & \!\!\! K{_{ab}}^{i}K^{abj}K_{cdj}K{^{cd}}_{k}\phi
^{k}- g(R(n_{j},e^{b},)n^{k},n^{i}) \phi^{j} \widetilde \nabla_{b}
K_{k} - \widetilde \nabla _{b}[g(R(n_{j},e^{b},)n^{k},n^{i}))
\phi^{j}K_{k}]
\nonumber \\
\!\!\! & + & \!\!\! K^{cdi}g(R(e_{c},n_{l})e_{d},n^{j})
\phi^{l}K_{j} + K^{cdj}g(R(e_{c},n_{l})e_{d},n^{i}) \phi^{l}K_{j}
+ K^{cdj}K{_{cd}}^{i}g(R(e_{a},n_{l})e^{a},n_{j}) \phi^{l} \nonumber \\
\!\!\! & - & \!\!\! K_{j}K^{i}g(R(e_{a},n_{l})e^{a},n^{j})
\phi^{l} - \frac{1}{2}g(R(e_{a},n_{j})e^{a},n^{i})
\phi^{j}K_{l}K^{l}  + g(R(e_{a},n^{l})e^{a},n^{i})K_{cdl}K^{cdj}
\phi_{j} \nonumber \\
 \!\!\! & + &  \!\!\! \widetilde
\Delta [g(R(e_{a},n^{j})e^{a},n^{i})] \phi_{j} +
2g(R(e_{a},n^{j})e^{a},n^{i}) \widetilde \Delta \phi_{j}
 + 2 \widetilde \nabla_{a}[g(R(e_{b},n^{j})e^{b},n^{i})]
\widetilde \nabla^{a} \phi_{j} \nonumber \\
 \!\!\! & - & \!\!\!
g(R(e_{a},n^{l})e^{a},n^{i})g(R(e_{b},n_{j})e^{b},n_{l}) \phi^{j}
- K_{j}\delta [g(R(e_{b},n^{j})e^{b},n^{i})] =0,
\end{eqnarray}
where equations (4.6) and (4.16) of \cite {5} have been employed.\\
For simplicity we set $K^{i}=0$ (extremal surfaces) in the
linearized equations (12), then the equation is reduced to
\begin{equation}
-(P^{2}){^{i}}_{j}\phi ^{j}=0,
\end{equation}
where the operator $P{^{i}_{j}}$ is given by,
\begin{equation}
P{^{i}}_{j}= \left[\widetilde \triangle ^{i}_{j}+
K{_{ab}}^{i}K{^{ab}}_{j} - g(R(e_{b},n_{j})e^{b},n^{i}) \right].
\end{equation}
 We can see that the
equation (14), is equal to the operator of the linearized
equations for [DNG] action, equation (4), which describes the
deformations of extremal surfaces in a curved
background.\\
Writing (13) explicitly, we find that
\begin{eqnarray}
\nonumber
(P^{2}){^{i}}_{j}\phi ^{j} \!\!\! & = & \!\!\! \widetilde \triangle \widetilde \triangle \delta ^{i}_{j}\phi ^{j}+2K{_{ab}}^{i}K{^{ab}}_{j}\widetilde \triangle \phi ^{j}+2\tilde \nabla ^{c}\left[K{_{ab}}^{i}K{^{ab}}_{j}\right](\widetilde \nabla _{c}\phi ^{j})+\widetilde \triangle \left[K{_{ab}}^{i}K{^{ab}}_{j}\right]\phi ^{j} \\
\!\!\! & + & \!\!\! K{_{ab}}^{i}K^{abk}K_{cdk}K{^{cd}}_{j}\phi^{j}
- 2g(R(e_{b},n^{j})e^{b},n^{i}) \widetilde \Delta \phi_{j} -
\widetilde \Delta [g(R(e_{b},n^{j})e^{b},n^{i})] \phi_{j}
\nonumber
\\
\!\!\! & - & \!\!\! K_{abj}K^{abl}g(R(e_{b},n^{j})e^{b},n^{i})
\phi_{l} - K{^{cd}}_{j}K{_{cd}}^{i}g(R(e_{b},n_{l})e^{b},n^{j})
\phi^{l} \nonumber \\
\!\!\! & - & \!\!\! 2\widetilde \nabla^{c}
[g(R(e_{b},n^{j})e^{b},n^{i})] \widetilde \nabla_{c} \phi_{j} +
g(R(e_{b},n^{j})e^{b},n^{i})g(R(e_{b},n_{l})e^{b},n_{j})
\phi^{l}=0.
\end{eqnarray}
\ It is remarkable that the solutions of the perturbations about
an extremal surface  for [DNG] action, equation (4), continue to
being solutions for [QEC], equation (15), even existing a curved
background. This is a more general result than that presented in
\cite{7}, for a flat spacetime.
\newline
\newline
\noindent \textbf{IV. Self-adjointness of the operators governing the deformations}\\[1ex]
 In this section, we shall show that the operator $(P^{2}){^{i}}_{j}$ in the
equation (14) is self-adjoint, which guarantees that a symplectic
current can be constructed in terms of solutions of equation (15).
For beginning, following \cite{13} we define $
M{^{i}}_{j}=M{_{j}}^{i}\equiv-K{_{ab}}^{i}K{^{ab}}_{j} +
g(R(e_{b},n_{j})e^{b},n^{i})$ as the mass matrix, therefore we can
rewrite the equation (15) as follows
\begin{eqnarray}
\nonumber
[\widetilde \triangle \widetilde \triangle \delta ^{i}_{j} \!\!\! & - & \!\!\! 2M{^{i}}_{j}\widetilde \triangle -2\tilde \nabla ^{c}[M{^{i}}_{j}](\widetilde \nabla _{c})-\widetilde \triangle [M{^{i}}_{j}] \\
\!\!\! & + & \!\!\! M^{ik}M_{kj}]\phi ^{j}=0.
\end{eqnarray}
Now, let $\phi ^{i}_{1}$ and $\phi ^{i}_{2}$ be two arbitrary
scalar fields, which correspond to a pair of solutions of equation
(15), thus is very easy to prove the following,
\begin{equation}
\phi_{1i}\widetilde \triangle \widetilde \triangle \phi
^{i}_{2}\equiv (\widetilde \triangle \widetilde \triangle \phi
_{1i})\phi ^{i}_{2}+ \nabla _{a}j{_{1}}^{a},
\end{equation}
where $j_{1}{^{a}}$ is given by
\begin{equation}
j_{1}{^{a}}=\phi _{1i}\widetilde \nabla ^{a}\widetilde \triangle
\phi _{2}^{i}+ \widetilde \triangle \phi _{1i} \widetilde \nabla
^{a}\phi _{2}^{i}-\widetilde \nabla ^{a}\phi _{1i}\widetilde
\triangle \phi _{2}^{i}-\widetilde \nabla ^{a}\widetilde \triangle
\phi _{1i}\phi _{2}^{i}.
\end{equation}
Furthermore,  we can demonstrate that,
\begin{equation}
-2M{^{i}}_{j}(\phi _{1i}\widetilde \triangle\phi _{2}^{j}) -2
\widetilde \nabla _{a}M{^{i}}_{j}(\phi _{1i}\widetilde \nabla
^{a}\phi _{2}^{j})\equiv -2M{^{i}}_{j}(\widetilde \triangle \phi
_{1i})\phi _{2}^{j}- 2\widetilde \nabla _{a}M{^{i}}_{j}(\widetilde
\nabla ^{a}\phi _{1i})\phi ^{j}_{2} + \nabla _{a}j_{2}^{a},
\end{equation}
with $j_{2}{^{a}}$:
\begin{equation}
j_{2}{^{a}}=-2\left[M{^{i}}_{j}\phi _{1i}\widetilde \nabla
^{a}\phi _{2}^{j}-M{^{i}}_{j}\widetilde \nabla ^{a}\phi _{1i}\phi
_{2}^{j}\right],
\end{equation}
finally we obtain, putting (18) and (20) together
\begin{eqnarray}
\nonumber
 \!\!\! & \phi _{1i} & \!\!\![\widetilde \triangle \widetilde \triangle \delta ^{ij} - 2M^{ij}\widetilde \triangle - 2\tilde \nabla ^{c}[M^{ij}](\widetilde \nabla _{c}) - \widetilde \triangle [M^{ij}] + M^{ik}M^{kj}]\phi_{2j} \\
 \!\!\! & = & \!\!\! [[\widetilde \triangle \widetilde \triangle \delta ^{ji} - 2M^{ji}\widetilde \triangle - 2\tilde \nabla ^{c}[M^{ji}](\widetilde \nabla _{c})-\widetilde \triangle [M^{ji}] + M^{jk}M^{ki}]\phi _{1i}]\phi_{2j} \nonumber \\
\!\!\! &  + & \!\!\!  \nabla _{a}j^{a},
\end{eqnarray}
where we have considered the symmetry of the mass matrix, and
$j^{a}$ is given by
\begin{eqnarray}
\nonumber j^{a} \!\!\! & = & \!\!\! \phi _{1i}\widetilde \nabla
^{a}\widetilde \triangle \phi _{2}^{i}+ \widetilde \triangle \phi
_{1i} \widetilde \nabla ^{a}\phi _{2}^{i}-\widetilde \nabla
^{a}\phi _{1i}\widetilde \triangle \phi _{2}^{i}-\widetilde \nabla
^{a}\widetilde \triangle \phi _{1i}\phi _{2}^{i} \\  \!\!\! & + &
\!\!\! 2K{_{ab}}^{i}K{^{ab}}_{j}\phi _{1i}\widetilde \nabla
^{a}\phi _{2}^{j} -2 g(R(e_{b},n_{j})e^{b},n^{i})\phi
_{1i}\widetilde \nabla ^{a}\phi _{2}^{j} \nonumber \\
\!\!\! & - & \!\!\!  2K{_{ab}}^{i}K{^{ab}}_{j}\widetilde \nabla
^{a}\phi _{1i}\phi _{2}^{j}+ 2g(R(e_{b},n_{j})e^{b},n^{i})
\widetilde \nabla ^{a}\phi _{1i}\phi _{2}^{j},
\end{eqnarray}
it is remarkable to see that there exist background gravity terms
in this expression for our symplectic current.\\
 Considering that $\phi _{1i}$ and $\phi _{2j}$ correspond to a
pair of solutions of the equation (15), equation (21) implies that
the operator $(P^{2}){^{i}}_{j}$ is self-adjoint, therefore, we
have
\begin{equation}
 \bigtriangledown
_{a}j^{a}=0.
\end{equation}
In the next section, we will take equation (22) on the phase space
for [QEC] theory and we will compared with the variation of
symplectic potential given in (11).
\newline
\newline
\noindent \textbf{V.  The Witten covariant phase space and the Symplectic Structure on Z}\\[1ex]
The basic idea of the covariant description of the canonical
formalism is to construct a symplectic structure on the classical
phase space, instead of choosing $p's$ and $q's$. In this manner,
in agreement with \cite{1}, the Witten phase space for [DNG-G]
theory is the space of solutions of equation (10), that we shall
call $Z$, and on such phase space we will construct a symplectic
structure. Thus, we can identify $e_{a}$, $n^{i}$, $k{_{ab}}^{i}$,
$\gamma_{ab}$ as zero-forms on $Z$, and the scalar fields
$\phi^{i}$ are closed one-forms on $Z$ (see Appendix, section IV),
it is
\begin{equation}
\widetilde D_{\delta} \phi^{i}=0.
\end{equation}
In this manner, considering the last paragraph, we can see that
the expression (22) is a covariantly conserved two-form on $Z$.
Thus, on the phase space $Z$ it is enough take only one solution
\cite{1}, then we can set $\phi_{1i}=\phi_{2i}=\phi_{i}$ in (22),
and we obtain without loosing generality
\begin{eqnarray}
\nonumber j^{a} \!\!\! & = & \!\!\! \phi _{i}\widetilde \nabla
^{a}\widetilde \triangle \phi^{i}+ \widetilde \triangle \phi _{i}
\widetilde \nabla ^{a}\phi^{i}-\widetilde \nabla ^{a}\phi
_{i}\widetilde \triangle \phi^{i}-\widetilde \nabla ^{a}\widetilde
\triangle \phi _{1i}\phi^{i} \\  \!\!\! & + & \!\!\!
2K{_{ab}}^{i}K{^{ab}}_{j}\phi _{i}\widetilde \nabla ^{a}\phi^{j}
-2 g(R(e_{b},n_{j})e^{b},n^{i})\phi
_{i}\widetilde \nabla ^{a}\phi^{j} \nonumber \\
\!\!\! & - & \!\!\!  2K{_{ab}}^{i}K{^{ab}}_{j}\widetilde \nabla
^{a}\phi _{i}\phi^{j}+ 2g(R(e_{b},n_{j})e^{b},n^{i}) \widetilde
\nabla ^{a}\phi _{i}\phi^{j},
\end{eqnarray}
considering that $\phi_{i}$ are one-forms on $Z$ and ( hence
$\widetilde \triangle \phi _{i}$, $\widetilde \nabla
^{a}\widetilde \triangle \phi^{i}$, $ \widetilde \nabla
^{a}\phi_{i}$), we have for example $(\widetilde \nabla
^{a}\phi^{i}) \phi_{i}=-\phi_{i}(\widetilde \nabla ^{a}\phi^{i})$,
and $j^{a}$ becomes to be
\begin{equation}
j^{a}= \phi_{i}\widetilde \nabla ^{a} \widetilde \Delta \phi^{i} +
\widetilde \Delta \phi^{i} \widetilde \nabla ^{a} \phi_{i} +
2K{_{cd}}^{{i}}K^{cdj}\phi_{i}\widetilde \nabla^{a}\phi_{j}-
2g(R(e_{b},n^{j})e^{b},n^{i})\phi_{i} \widetilde
\nabla^{a}\phi_{j},
\end{equation}
that we will use in this section. Strictly $\widetilde
\nabla^{a}\phi^{i} \phi_{i}$ corresponds to the wedge product of
one-forms on $Z$, $\widetilde \nabla^{a}\phi^{i} \wedge \phi_{i}$,
but in this paper we omit the
explicit use of $\wedge$ [see for example \cite{1}].\\
On the other hand, the symplectic structure on Z is a
(non-degenerate) closed two-form; the closeness is equivalent to
the Jacobi identity in the conventional Hamiltonian scheme, and
the antisymmetry of a two-form represents the antisymmetry of
Poisson brackets. In this section, we will find a covariant
symplectic structure for [QEC] theory, and we will demonstrate
that such a geometric structure is even an exact two-form (which
implies that in particular
is closed). \\
To prove the closeness, we shall calculate the variation of
$\Psi^{a}$. For beginning, we will calculate for an arbitrary
field $\psi^{i}$ the variation $\widetilde{D}_{\delta}
\widetilde{\nabla}_{b} \psi^{i}$, this is \cite{5},
\begin{eqnarray}
     \widetilde{D}_{\delta} \widetilde{\nabla}_{b} \psi^{i} \!\! & = &
\!\! D_{\delta} [D_{b} \psi^{i} - \omega_{b} {^{ij}} \psi_{j}] -
\gamma^{ij} \widetilde{\nabla}_{b} \psi_{j} \\
     \!\! & = & \!\! D_{b} D_{\delta} \psi^{i} - (D_{\delta} \omega_{b}
{^{ij}})\psi_{j} - \omega_{b} {^{ij}} D_{\delta} \psi_{j} -
\widetilde{\nabla}_{b} (\gamma^{ij} \psi_{j}) +
(\widetilde{\nabla}_{b} \gamma^{ij}) \psi_{j} \nonumber \\
     \!\! & = & \!\! \widetilde{\nabla}_{b} \widetilde{D}_{\delta}
\psi^{i} - (D_{\delta} \omega_{b} {^{ij}} - \widetilde{\nabla}_{b}
\gamma^{ij}) \psi_{j}. \nonumber
\end{eqnarray}
Using the equations (24) and (27), we have for
$\psi^{i}=\phi^{i}$,
\begin{eqnarray}
\widetilde{D}_{\delta} \widetilde{\nabla}_{b} \phi^{i} \!\! & = &
\!\! -(D_{\delta} \omega_{b} {^{ij}} - \widetilde{\nabla}_{b}
\gamma^{ij}) \phi_{j} \nonumber \\
\!\! & = & \!\! K_{bc}{^{i}}\widetilde\nabla^{c}
\phi^{j}\phi_{j}-K_{bc}{^{j}}\widetilde
\nabla^{c}\phi^{i}\phi_{j}-g(R(n_{k},e_{b})n^{j},n^{i})\phi^{k}\phi_{j},
\end{eqnarray}
where we have used that \cite{5}
\begin{equation}
D_{\delta} \omega_{a} {^{ij}} - \widetilde{\nabla}_{a} \gamma^{ij}
= - K_{ab} {^{i}} \widetilde{\nabla}^{b} \phi^{j} + K_{ab} {^{j}}
\widetilde{\nabla}^{b} \phi^{i}+ g(R(n_{k}, e_{a}) n^{j}, n^{i})
\phi^{k}.
\end{equation}
Similarly, using the equation (27), (29) and $\widetilde
D_{\delta}K^{i}=-\widetilde
\triangle\phi^{i}-K{_{ab}}^{i}K^{ab}{_{j}}\phi^{j}+g(R(e_{a},n_{j}),e^{a},n^{i})\phi^{j}$
(see \cite{5}), we obtain for $\psi^{i}=K^{i}$,
\begin{eqnarray}
\nonumber \widetilde{D}_{\delta} \widetilde{\nabla}_{b} K^{i} \!\!
& = & \!\! \widetilde{\nabla}_{b} \widetilde{D}_{\delta} K^{i} -
(D_{\delta} \omega_{b} {^{ij}} - \widetilde{\nabla}_{b}
\gamma^{ij}) K_{j}\\
\!\! & = & \!\! -\widetilde \nabla_{b}\widetilde \Delta \phi^{i} - K_{cd}{^{i}}K^{cdj}\widetilde \nabla_{b}\phi_{j} - \widetilde \nabla_{b}(K_{cd}{^{i}}K^{cdj})\phi_{j} + \widetilde \nabla_{b}g(R(e_{a},n^{j})e^{a},n^{i})\phi_{j} \nonumber \\
\!\! & + & \!\! g(R(e_{a},n^{j})e^{a},n^{i})\widetilde
\nabla_{b}\phi_{j}+ K_{bc}{^{i}}\widetilde\nabla^{c}
\phi^{j}K_{j}-K_{bc}{^{j}}\widetilde
\nabla^{c}\phi^{i}K_{j}-g(R(n_{k},e_{b})n^{j},n^{i})\phi^{k}K_{j}.
\end{eqnarray}
On the other hand, rewriting the symplectic potential as
$\Psi^{b}= \sqrt{-\gamma}h^{b}$, with $h^{b}=\frac{1}{2}K^{j}K_{j}
\phi^{b} + \phi_{i} \widetilde \nabla^{b}K^{i}-K_{i} \widetilde
\nabla^{b}\phi^{i}$, it is easy to verify that,\\
\begin{eqnarray}
D_{\delta}\Psi_{b} \!\! & = & \!\! \widetilde D_{\delta}
(\sqrt{-\gamma} h_{b}) \nonumber \\  \!\! & = & \left[\!\!
\widetilde D_{\delta} \sqrt{-\gamma} \right] h_{b}
\nonumber \\
\!\! & + & \!\! \sqrt{-\gamma} \left[\widetilde D_{\delta}(\frac{1}{2}K^{j}K_{j} \phi_{b}) +  \widetilde D_{\delta}\widetilde \nabla_{b}K^{i}\phi_{i} + \widetilde \nabla_{b}K_{i} \widetilde D_{\delta}\phi^{i} -\widetilde D_{\delta}K^{i} \widetilde \nabla_{b} \phi_{i} -K_{i} \widetilde D_{\delta} \widetilde \nabla_{b}\phi^{i} \right] \nonumber \\
\!\! & = & \!\! \sqrt{-\gamma}
[(K_{i}K^{i}K^{j}\phi_{j}+K_{i}(-\widetilde \triangle
\phi^{i}+g(R(e_{a},n_{j})e^{a},n^{i}) \phi
^{j}-K{^{ab}}^{i}K_{abj}\phi ^{j}))\phi_{b} \nonumber \\
\!\!\ & + & \!\!\  \frac{1}{2}K^{j}K_{j} K{_{bc}}^{i} \phi_{i}
\phi^{c}- \frac{1}{2}K^{j}K_{j} \phi_{i} \widetilde \nabla _{b}
\phi^{i}  \nonumber \\
\!\! & - & \!\!  K^{j}\phi_{j}K_{i}\widetilde \nabla_{b}\phi^{i} - \widetilde \nabla _{b} \widetilde \Delta \phi^{i} \phi_{i}+2K_{cd}{^{i}K^{cdj}\phi_{i}\widetilde \nabla_{b}\phi_{j}}+  K^{j}\phi_{j}\phi_{i}\widetilde \nabla_{b}K^{i}  \nonumber \\
\!\! & - & \!\! 2g(R(e_{a},n^{j})e^{a},n^{i})\phi_{i}\widetilde \nabla_{b}\phi_{j} -2K_{bc}{^{j}}\phi_{j}\widetilde  \nabla ^{c}\phi^{i}K_{i} + \widetilde \Delta \phi^{i} \widetilde \nabla _{b} \phi_{i} \nonumber \\
\!\! & - & \!\! 2 K_{i}K_{bc}{^{i}}\widetilde \nabla
^{c}\phi^{j}\phi_{j} -
2g(R(n_{k},e_{b})n^{j},n^{i})\phi^{k}K_{j}\phi_{i}],
\end{eqnarray}
where we have used the equations (24), (28) and (30). In this
manner, and in concordance with equation (31), we can see that
\begin{eqnarray}
\nonumber \delta\Psi^{a} \!\!\ & = & \!\!\
D_{\delta}\Psi^{a}=\widetilde D_{\delta}(\gamma^{ab}\Psi_{b})
\\ \!\! & = & \!\! \widetilde
D_{\delta}\gamma^{ab}\Psi_{b}+\gamma^{ab}\widetilde
D_{\delta}\Psi_{b}
 =\sqrt{-\gamma}j'^{a},
\end{eqnarray}
where
\begin{eqnarray}
\nonumber j'^{a} \!\! & = & \!\!
(K_{i}K^{i}K^{j}\phi_{j}+K_{i}(-\widetilde \triangle
\phi^{i}+g(R(e_{b},n_{j})e^{b},n^{i}) \phi
^{j}-K{^{cd}}^{i}K_{cdj}\phi ^{j}))\phi^{a} \nonumber \\
\!\!\ & - & \!\!\ \frac{1}{2}K^{j}K_{j}K^{aci} \phi_{i} \phi_{c} -
\frac{1}{2} K^{j}K_{j} \phi_{i} \widetilde \nabla^{a} \phi^{i}
\nonumber \\
\!\! & + & \!\!  \phi_{i}\widetilde \nabla ^{a} \widetilde \Delta \phi^{i} + \widetilde \Delta \phi^{i} \widetilde \nabla ^{a} \phi_{i} + 2K{_{cd}}^{{i}}K^{cdj}\phi_{i}\widetilde \nabla^{a}\phi_{j} \nonumber \\
\!\! & - & \!\! 2g(R(e_{b},n^{j})e^{b},n^{i})\phi_{i} \widetilde \nabla^{a}\phi_{j} -2g(R(n_{k},e^{a})n^{j},n^{i})\phi^{k}K_{j}\phi_{i} - 2K^{ab}{^{j}}\phi_{j}\widetilde \nabla_{b}K^{i}\phi_{i} \nonumber \\
\!\! & - & \!\! 2 K_{i}K^{ab}{^{i}}\widetilde \nabla
_{b}\phi^{j}\phi_{j} + K^{j}\phi_{j}\phi_{i}\widetilde
\nabla_{b}K^{i} - K^{j}\phi_{j}K_{i}\widetilde \nabla_{b}\phi^{i},
\end{eqnarray}
considering again that $\phi^{a}$ is a diffeomorphism on the
worldvolume and therefore it can be gauged away $(\phi^{a}=0)$, we
obtain finally
\begin{eqnarray}
\nonumber j'^{a} \!\! & = & \!\!  - \frac{1}{2} K^{j}K_{j}
\phi_{i} \widetilde \nabla^{a} \phi^{i}
 +   \phi_{i}\widetilde \nabla ^{a} \widetilde \Delta \phi^{i} + \widetilde \Delta \phi^{i} \widetilde \nabla ^{a} \phi_{i} + 2K{_{cd}}^{{i}}K^{cdj}\phi_{i}\widetilde \nabla^{a}\phi_{j} \nonumber \nonumber \\
\!\! & - & \!\! 2g(R(e_{b},n^{j})e^{b},n^{i})\phi_{i} \widetilde \nabla^{a}\phi_{j} -2g(R(n_{k},e^{a})n^{j},n^{i})\phi^{k}K_{j}\phi_{i} - 2K^{ab}{^{j}}\phi_{j}\widetilde \nabla_{b}K^{i}\phi_{i} \nonumber \\
\!\! & - & \!\! 2 K_{i}K^{ab}{^{i}}\widetilde \nabla
_{b}\phi^{j}\phi_{j} + K^{j}\phi_{j}\phi_{i}\widetilde
\nabla_{b}K^{i} - K^{j}\phi_{j}K_{i}\widetilde \nabla_{b}\phi^{i}.
\end{eqnarray}
If we take the special case of  extremal surfaces $(K^{i}=0)$ in a
curved background, we find
\begin{equation}
j'^{a}= \phi_{i}\widetilde \nabla ^{a} \widetilde \Delta \phi^{i}
+ \widetilde \Delta \phi^{i} \widetilde \nabla ^{a} \phi_{i} +
2K{_{cd}}^{{i}}K^{cdj}\phi_{i}\widetilde \nabla^{a}\phi_{j}-
2g(R(e_{b},n^{j})e^{b},n^{i})\phi_{i} \widetilde
\nabla^{a}\phi_{j},
\end{equation}
which corresponds exactly to the current found in the Section III
(equation (26)), using the self-adjoint operators method. \\
With these results,  we can define a two-form in terms of
$j'^{a}$, given in (35), that will be the symplectic structure
that we require,
\begin{equation}
\omega\equiv \int_\Sigma \sqrt{-\gamma}j'^{a}d\Sigma_{a}.
\end{equation}
where $\Sigma$ is a Cauchy p-surface.\\
In this manner, we can see that our symplectic structure is an
exact two-form because it comes from a exterior derivative of the
global symplectic potential on $Z$ (equation (32)) and it is in
particular closed due to that $\delta$ is nilpotent, this is
\begin{equation}
\delta \omega=\int_\Sigma \delta(\delta\Psi^{a})d\Sigma_{a}=0,
\end{equation}
therefore, we can see that it is advantageous to identify a
symplectic potential from total divergences terms, that in the
literature are neglected, but it is relevant on the phase space
$Z$, since
it generates our symplectic structure $\omega$ by means of a direct exterior derivative.  \\
On the other hand, because of $\nabla_{a}j´^{a}=0$, we can use the
Stokes theorem in (36) and see that $\omega$ is independent on the
choice of $\Sigma$, this is $\omega_{\Sigma}= \omega_{\Sigma'}$.
It will be a very important property of $\omega$, since it allows
us to establish a connection between functions and
Hamiltonian vector fields on $Z$; this subject will be considered in the future works.\\
Now, we shall prove that the symplectic structure that we have
found (equation (36)) is invariant under infinitesimal spacetime
diffeomorphism, which corresponds to the gauge directions of the
theory on $Z$ \cite{1, 4, 6}. For this purpose, let us consider
first the indeterminacy directions of the global symplectic
potential on $Z$, and we notice from equation (32) that there are
more than one symplectic potential, since $\delta$ is nilpotent,
this is
\begin{equation}
\sqrt{-\gamma}j'^{a}= \delta(\Psi^{a} + \delta \eta^{a}),
\end{equation}
where $\eta^{a}$ is an arbitrary (worldvolume) field. On
the other hand, we know that an infinitesimal spacetime
diffeomorphism $(\delta X^{\mu})$ induces a infinitesimal
diffeomorphism on the worldvolume $(\delta \xi^{a}) $, this is
\begin{equation}
\delta X^{\mu}=\epsilon^{a} \partial_{a} X^{\mu}=e_{a}^{\mu}
\delta\xi^{a},
\end{equation}
where $e_{a}^{\mu}=\frac{ \partial X^{\mu}}{ \partial\xi^{a}}$ and
$\xi^{a}$ are the world-volume coordinates. In this manner, in
(38) we can identify that, in particular, $\eta^{a}= \xi^{a}$,
then the indeterminacy directions of the symplectic potential,
contain, in particular, the gauge directions
of the theory, therefore we have showed that $\omega$ is a no-degenerate two-form on $Z$.\\
Furthermore, since $Z$ is the set of solutions of equations of
motion, it let $\widehat{Z}$ be the space of solutions modulo
gauge transformations or quotient space $\widehat{Z}=Z/G$, where
$G$ is the group of spacetime diffeomorphisms; then we have that
$j'^{a}$ has vanishing components along the $G$ orbits and
therefore $\omega$ too.
\newline
\newline
\newline
\noindent \textbf{VIII. Conclusions and prospects}\\[1ex]
As we have seen, the arguments of total divergences for the
theories under study are identified as  symplectic potentials,
that are not relevant in the dynamics of the system, but are very
important in the corresponding Witten covariant phase space, since
they generate geometrical structures for [DNG] and [QEC] theories,
confirming the results previously obtained in \cite{4} for the
former, and creating a symplectic structure  for the later, that
we will use in the future for identify the canonically conjugate
variables, construct, for example, the corresponding Poisson
brackets,  find relevant symmetries, and study the quantization
aspects for [QEC] theory.
[QEC].\\
It is important to mention that the treatment in this paper for
[QEC] p-branes is general, and contains the particular case of
[QEC] string, in this manner, we have the necessary elements for
study the quantization aspects of a different system (in this case
[QEC] string)  to that we commonly find in the literature, namely
[DNG] string. For this aim, we need the symplectic structure that
we have constructed, and solve the equations of motion, equation
(10), which is crucial in the study of such
aspects.\\
As we know, to solve the equations of motion, (10), for p-branes
is very difficult; however, $K^{i}=0$ is a subset of solutions of
such equations, and in the literature we find the solutions to
$K^{i}=0$ corresponding to extremal surfaces, for [DNG] string,
that are well know. In this manner, taking as particular case
[QCE] string, we can use the same solutions to the equations of
motion for [DNG] string, as a subset of solutions to equations
(10) and the symplectic structure that we have constructed,
equation (36), to study in an explicit way the quantization
aspects for
[QCE] strings, leaving it as a future work.\\
In addition, as a future work, we will show that finding a
symplectic potential as in the presented  work, we can identify
the contributions of the  topological terms in a canonical scheme,
 which is completely unknown in the
literature \cite{14}.\\[1ex]

\noindent \textbf{Acknowledgements}\\[1ex]
This work was supported by CONACYT. The author wants to thank R.
Cartas-Fuentevilla for drawing my interest to the study of branes.
\\[1em]

\noindent {\uno  Appendix} \vspace{1em}\\
\noindent \textbf{ GEOMETRY OF THE EMBEDDING AND THEIR DEFORMATIONS}\\[1ex]
\noindent \textbf{ I. The embedding}\\[1ex]
The $D$-dimensional brane dynamics is usually given by a oriented
timelike worldsheet {\it m} described by the embedding functions
$x^{\mu} = X^{\mu} (\xi^{a})$, $\mu = 0, ..., N - 1$ and $ a = 0,
..., D$, in a $N$-dimensional ambient spacetime {\it M} endowed
with the metric $g_{\mu\nu}$. Such functions specify the
coordinates of the brane, and the $\xi^{a}$ correspond to internal
coordinates on the worldsheet.

At each point of {\it m}, $e_{a} \equiv X^{\mu}_{,a}
\partial_{\mu} \equiv e^{\mu}_{a} \partial_{\mu}$, generate a basis
of tangent vectors to {\it m}; thus, the induced
$(D+1)$-dimensional worldsheet metric is given by $\gamma_{ab} =
e^{\mu}_{a} e^{\nu}_{b} g_{\mu\nu} = g(e_{a}, e_{b})$.
Furthermore, the $(N - D)$ vector fields $n^{i}$ normal to {\it
m}, are defined by
\begin{equation}
     g(n^{i}, n^{j}) = \delta^{ij}, \quad g(e_{a}, n^{i}) = 0.
\end{equation}
Tangential indices are raised and lowered by $\gamma^{ab}$ and
$\gamma_{ab}$, respectively, whereas normal vielbein indices by
$\delta^{ij}$ and $\delta_{ij}$ respectively, and this fact will
be used implicitly below. The collection of vectors $\{e_{a},
n^{i}\}$, which can be used as a basis for the spacetime vectors,
satisfies the generalized Gauss-Weingarten equation:
\[
   D_{a} e_{b} = \gamma_{ab} {^{c}}e_{c} - K_{ab} {^{i}}n_{i},
\quad D_{a} n_{i} = K_{ab} {^{i}}e^{b} + \omega_{a} {^{ij}}n_{j},
\]
where $D_{a} \equiv e^{\mu}_{a} D_{\mu}$ ($D_{\mu}$ is the
torsionless covariant derivative associated with $g_{\mu\nu}$);
thus, the connection coefficients $\gamma_{ab} {^{a}}$ compatible
with $\gamma_{ab}$ is given by $\gamma_{ab} {^{c}} = g(D_{a}
e_{b}, e^{c}) = \gamma_{ba} {^{c}}$, and the {\it i}th extrinsic
curvature of the worldsheet  by $K_{ab} {^{i}} = - g (D_{a} e_{b},
n^{i}) = K_{ba}{^{i}}$. Similarly the extrinsic twist potential of
the worldsheet is defined by $\omega_{a} {^{ij}} = g(D_{a} n^{i},
n^{j}) = - \omega_{a} {^{ji}}$. Such a potential allows us to
introduce a worldsheet covariant derivative
$(\widetilde{\nabla}_{a})$ defined on fields $(\Phi^{i} {_{j}})$
transforming as tensors under normal frame rotations:
\begin{equation}
     \widetilde{\nabla}_{a} \Phi^{i} {_{j}} \equiv \nabla_{a}
\Phi^{i} {_{j}} - \omega_{a} {^{ik}} \Phi_{k j} - \omega_{ajk}
\Phi^{ik},
\end{equation}
where $\nabla_{a}$ is the (torsionless) covariant derivative
associated with $\gamma_{ab}$. \\[2em]

\noindent \textbf{II. Deformations of the intrinsic geometry}\\[1ex]
The deformation of the embedding is given by an arbitrary
infinitesimal deformation $\delta X^{\mu}$, decompose  into its
parts tangential and normal to the worldsheet
\begin{equation}
\delta X^{\mu}=e{_{a}}^{\mu} \phi^{a} + n{_{i}}^{\mu} \phi^{i},
\end{equation}
where
\begin{equation}
\phi^{a}= \delta X^{\mu}e{^{a}}_{\mu} \quad {\rm{and}} \quad
\phi^{i}= \delta X^{\mu} n{^{i}}_{\mu},
\end{equation}
however, in this scheme of deformations \cite{5}, the physically
observable measure of the deformation of the embedding {\it m}, is
given by the orthogonal projection of the infinitesimal spacetime
variation $\xi^{\mu} \equiv \delta X^{\mu} = n^{\mu}_{i}
\phi^{i}$, characterized by $N-D$ scalar fields $\phi^{i}$, and
the scalar fields $\phi^{a}= \delta X^{\mu}e{^{a}}_{\mu}$, are
neglected because of it is identify as a diffeomorphism on the
worldsheet.\\
Defining the vector field $\delta \equiv n_{i} \phi^{i}$, the
displacement induced in the tangent basis $\{ e_{a} \}$ along
$\delta$ depends on $\phi^{i}$ and on their first derivatives:
\begin{equation}
     D_{\delta} e_{a} = \beta_{ab} e^{b} + J_{ai}n^{i},
\end{equation}
where $D_{\delta} \equiv \delta^{\mu} D_{\mu}$, and
\begin{equation}
     \beta_{ab} = g(D_{\delta} e_{a}, e_{b}) = K_{ab} {^{i}}
\phi_{i}, \quad J_{ai} = g (D_{\delta} e_{a}, n_{i}) =
\widetilde{\nabla}_{a} \phi_{i};
\end{equation}
similarly, the deformation in the induced metric on {\it m} is
given by
\begin{equation}
     D_{\delta} \gamma_{ab} = 2 \beta_{ab} = 2 K_{ab} {^{i}}
\phi_{i}, \quad D_{\delta} \gamma^{ab} = - 2 \beta^{ab}.
\end{equation}
For the case treated here, this is sufficient about the
deformations of the intrinsic geometry. \\[2em]

\noindent \textbf{III. Deformations of the extrinsic geometry}\\[1ex]
Introducing a covariant deformation derivative as
$\widetilde{D}_{\delta} \Psi_{i} \equiv D_{\delta} \Psi_{i} -
\gamma_{i}{^{j}} \Psi_{j}$, where $\gamma_{ij} = g(D_{\delta}
n_{i}, n_{j}) = - \gamma_{ij}$, the covariant measure of the
deformations of the quantities characterizing the extrinsic
geometry are given by
\begin{eqnarray}
 \!\! & & \!\! D_{s} n_{i} = - J_{ai} e^{a} + \gamma_{ij}
n^{j}, \quad \widetilde{D}_{\delta} n_{i} = - J_{ai} e^{a} =
- (\widetilde{\nabla}_{a} \phi_{i}) e^{a}, \\
     \!\! & & \!\! \widetilde{D}_{\delta} K_{ab} {^{i}} =
- \widetilde{\nabla}_{a} \widetilde{\nabla}_{b} \phi^{i} + [K_{ac}
{^{i}} K^{c} {_{bj}}-g(R(e_{a}, n_{j}) e_{b}, n^{i})]
\phi^{j}, \\
     \!\! & & \!\! \widetilde{D}_{\delta} \omega_{a} {^{ij}}
- \nabla_{a} \gamma^{ij} =  D_{\delta} \omega_{a} {^{ij}} -
\widetilde{\nabla}_{a} \gamma^{ij} = - K_{ab} {^{i}}
\widetilde{\nabla}^{b} \phi^{j} + K_{ab} {^{j}}
\widetilde{\nabla}^{b} \phi^{i}  \nonumber \\
     \!\! & & \!\! \hspace{6.5cm} + g(R(n_{k}, e_{a}) n^{j}, n^{i})
\phi^{k},
\end{eqnarray}
which depend on second derivatives of $\phi_{i}$; the notation
$g(R(Y_{1}, Y_{2})Y_{3}, Y_{4}) = R_{\mu\nu\alpha\beta}
Y^{\nu}_{1} Y^{\mu}_{2} Y^{\alpha}_{3} Y^{\beta}_{4}$ is used,
where $R_{\mu\nu\alpha\beta}$ is the Riemann tensor of spacetime.
Other useful formulae and more details can be found directly in
Ref.\ \cite{5}. \\[2em]
\noindent \textbf{IV. The exterior calculus on Z}\\[1ex]
The space phase for [DNG] and [QEC] theories, is the space of
solutions of equations (4) and (10) respectively and we call $Z$
(see Section V). Any background quantity, such as those defined in
Section II of this appendix, will be associated with zero-forms on
$Z$ \cite{4}.\\
 On $Z$, the deformation operator $\delta$ acts as an
exterior derivative, taking $k$-forms into $(k+1)$-forms, and it
should satisfy
\begin{equation}
\delta^{2}=0,
\end{equation}
and the Leibniz rule
\begin{equation}
\delta(AB)=\delta A B- A \delta B.
\end{equation}
In this manner, $\delta X^{\mu}$ is the exterior derivative of the
zero-form $X^{\mu}$, and it be closed
\begin{equation}
\delta^{2} X^{\mu}=0.
\end{equation}
Thus, since $\phi^{i}= \delta X^{\mu} n{^{i}}_{\mu}$ and
$\phi^{a}= \delta X^{\mu}e{^{a}}_{\mu}$, corresponds to zero-forms
on $Z$, and there are anticommutating objets:
$\phi^{i}\phi^{j}=-\phi^{j}\phi^{i}$, $\phi^{a}\phi^{b}=-
\phi^{b}\phi^{a}$. In general, the differential forms satisfy the Grassman algebra, $AB= (-1)^{AB}BA$\\
It is remarkable to mention, that the covariant deformations
operator $D_{\delta}$ (and subsequently $\widetilde D_{\delta}$)
also works as an exterior derivative on $Z$, in the sense that
maps $k$-forms into ($k+1$)-forms; however $\widetilde
D_{\delta}^{2}$ does not vanish necessarily. In this manner, from
equation (44) we can identify $\beta_{ab}$ and $j_{ai}$ as one-
forms on $Z$.\\
Whit these preliminaries, it is very easy to prove the following
\cite {4}
\begin{equation}
D_{\delta}(\delta X^{\mu})=0,
\end{equation}
and
\begin{equation}
\widetilde D_{\delta}(\phi^{i})=0.
\end{equation}
Finally, using the equations (43), (44), (46), (51) and (53) we
can calculate the exterior derivative of $\phi^{a}$ on $Z$,
obtaining
\begin{eqnarray}
\nonumber \delta \phi^{a}= D_{\delta}( \gamma^{ab}\phi_{b}) \!\!\
& = & \!\!\
(D_{\delta}\gamma^{ab})\phi_{b}+\gamma^{ab}D_{\delta}(\phi_{b})
\nonumber \\
\!\!\ & = & \!\!\ -2 K^{abi} \phi_{i} \phi_{b}+
\gamma^{ab}[D_{\delta} (\delta X^{\mu})e_{b \mu}- \delta X^{\mu}
D_{\delta} e_{b \mu}] \nonumber \\
\!\!\ & = & \!\!\ -2 K^{abi} \phi_{i} \phi_{b} - \gamma^{ab}
\delta X^{\mu}[ K^{abi} \phi_{i} e_{b \mu} +
 \widetilde \nabla^{a}\phi^{i} n_{i \mu}] \nonumber \\
\!\!\ & = & \!\!\ -2 K^{abi} \phi_{i} \phi_{b} - \gamma^{ab}[
-K^{abi} \phi_{i} \phi_{b} -
 \widetilde \nabla^{a}\phi^{i} \phi_{i}]
 \nonumber \\  \!\!\ & =& \!\!\ - [K^{abi} \phi_{i}
\phi_{b} + \phi_{i} \widetilde \nabla^{a}\phi^{i} ].
\end{eqnarray}

\end{document}